\documentclass{iau} 
\usepackage{graphicx}
\usepackage{arydshln}

\title[Carbon-enhanced metal-poor 3D model atmospheres
] 
{Carbon-enhanced metal-poor 3D model atmospheres
}

\author[M. Steffen et al.]   
{M.~Steffen$^1$,
 A.~J.~Gallagher$^{2,3}$, E.~Caffau$^3$, P.~Bonifacio$^3$, and H.-G.~Ludwig$^4$}

\affiliation{$^1$Leibniz-Institut f\"ur Astrophysik Potsdam, 
D-14482 Potsdam, Germany \\[\affilskip] 

$^2$Max Planck Institut f\"ur Astronomie, D-69117 Heidelberg, Germany \\[\affilskip]
$^3$Observatoire de Paris, PSL Research University, CNRS, France\\[\affilskip]
$^4$LSW, Zentrum f\"ur Astronomie,  D-69120 Heidelberg, Germany
}

\pubyear{2017}
\volume{334}  
\jname{Rediscovering our Galaxy}
\editors{C. Chiappini, I. Minchev, E. Starkenburg, \& M. Valentini, eds.}
\begin{document}

\maketitle

\begin{abstract}
We present our latest 3D model atmospheres for carbon-enhanced metal-poor
(CEMP) stars computed with the CO5BOLD code. The stellar parameters are
representative of hot turn-off objects ($T_{\rm eff}\sim 6250$\,K, $\log g=4.0$,
[Fe/H]=3). The main purpose of these models is to investigate the role of 3D
effects on synthetic spectra of the CH G-band (4140-4400\,\AA), the CN 
BX-band (3870-3890\,\AA), and several UV OH transitions (3122-3128\,\AA). By
comparison with the synthetic spectra from standard 1D model atmospheres
(assuming local thermodynamic equilibrium, LTE), we derive 3D abundance 
corrections for carbon and oxygen of up to $-0.5$ and $-0.7$\,dex, respectively.
\end{abstract}

\firstsection 
\section{Introduction}
\emph{C, N, O in extremely metal-poor stars.}
Molecular bands of CH, CN and NH are the only available
indicators of the abundances of carbon and nitrogen in stars with
a metallicity below [M/H]=$-2.0$. Oxygen can be measured in giant
stars from the [OI]\,630\,nm line down to [M/H]\,$\approx$$-3.5$.
For turn-off and main sequence stars, only the OH lines are
available to get access to the oxygen abundance at these low
metallicities. For this reason, understanding the formation of the
molecular bands in extremely metal-poor stars is crucial for
deciphering the galactic chemical evolution of the key elements C, N,
and O.

\emph{The role of photospheric convection.}
The photospheres of metal-poor stars are known to be affected by
small-scale convective flows that give rise to horizontal temperature
fluctuations (“stellar granulation”) and extra surface cooling, effects
that cannot be captured by 1D hydrostatic radiative equilibrium model 
atmospheres. The strength of spectral lines forming in an inhomogeneous,
metal-poor stellar atmosphere tends to be enhanced with respect to the
prediction of standard 1D models. Spectroscopic analyses based on 
realistic 3D atmospheres therefore generally yield lower chemical abundances.

\section{3D model atmospheres}
For this study, we carried out 3D numerical simulations of convection in a
metal-poor stellar atmosphere with $T_{\rm eff} = 6250$\,K, $\log g = 4.0$, 
and metallicity [M/H]=$-3$, assuming three different sets of C, N, O 
abundances, as consistently reflected in the 14 bin opacity tables.  
Models {\bf B} and {\bf C} are carbon-enhanced by a factor $100$ but 
with different C/O ratio, while model {\bf A }has a ``normal'' metal-poor 
composition ([$\alpha$/Fe]=$0.4$). Details are given in Table\,\ref{tab1}.

\begin{table}
  \begin{center}
  \caption{\emph{Left:} C, N, O abundances adopted for the opacity tables of 
           3D models {\bf A}, {\bf B}, {\bf C}. \emph{Right:} 3D abundance 
           corrections derived from the comparison of 3D and 1D synthetic 
           spectra of the G-band and OH-band, respectively.}
  \label{tab1}
 {\scriptsize
  \begin{tabular}{l c c c c| @{\extracolsep{7pt}}l@{\extracolsep{0pt}} c c 
@{\extracolsep{10pt}}r@{\extracolsep{0pt}}}
\noalign{\smallskip}\hline\noalign{\smallskip}
Model & \multicolumn{4}{c}{Model abundances} & Synthesis & 
        \multicolumn{2}{c}{$\Delta_{\rm 3D-1D}$} & Comment \\ 
\cline{2-5} \cline{7-8}\noalign{\smallskip} 
ID    & $A$(C) & $A$(N) & $A$(O) & C/O & ID & G-band & OH-band \\ 
\hline\noalign{\smallskip}
{\bf A} & 5.39 & 4.78 & 6.06 & 0.21 & {\bf SAB} & $-0.76$ & $-0.40$& inconsistent \\ 
        &      &      &      &      & {\bf SAC} & $-0.31$ & $-0.88$& inconsistent \\ 
\hdashline\noalign{\smallskip}
{\bf B} & 7.39 & 6.78 & 6.06 & 21.4 & {\bf SB} & $-0.50$ & $-0.35$ & consistent\\
{\bf C} & 7.39 & 6.78 & 7.66 & 0.54 & {\bf SC} & $-0.32$ & $-0.73$ & consistent\\ \hline
  \end{tabular}
  }
 \end{center}
\vspace{1mm}
 \scriptsize{
 {\it Notes:}
  {\bf SAB} and {\bf SAB} indicate that the atmospheric
  structure of model {\bf A} was used together with (inconsistent)
  abundances {\bf B} and {\bf C}, respectively, for the spectrum synthesis.}
\end{table}

\begin{figure}[htb]
\begin{center}
 \includegraphics[width=6.7cm]{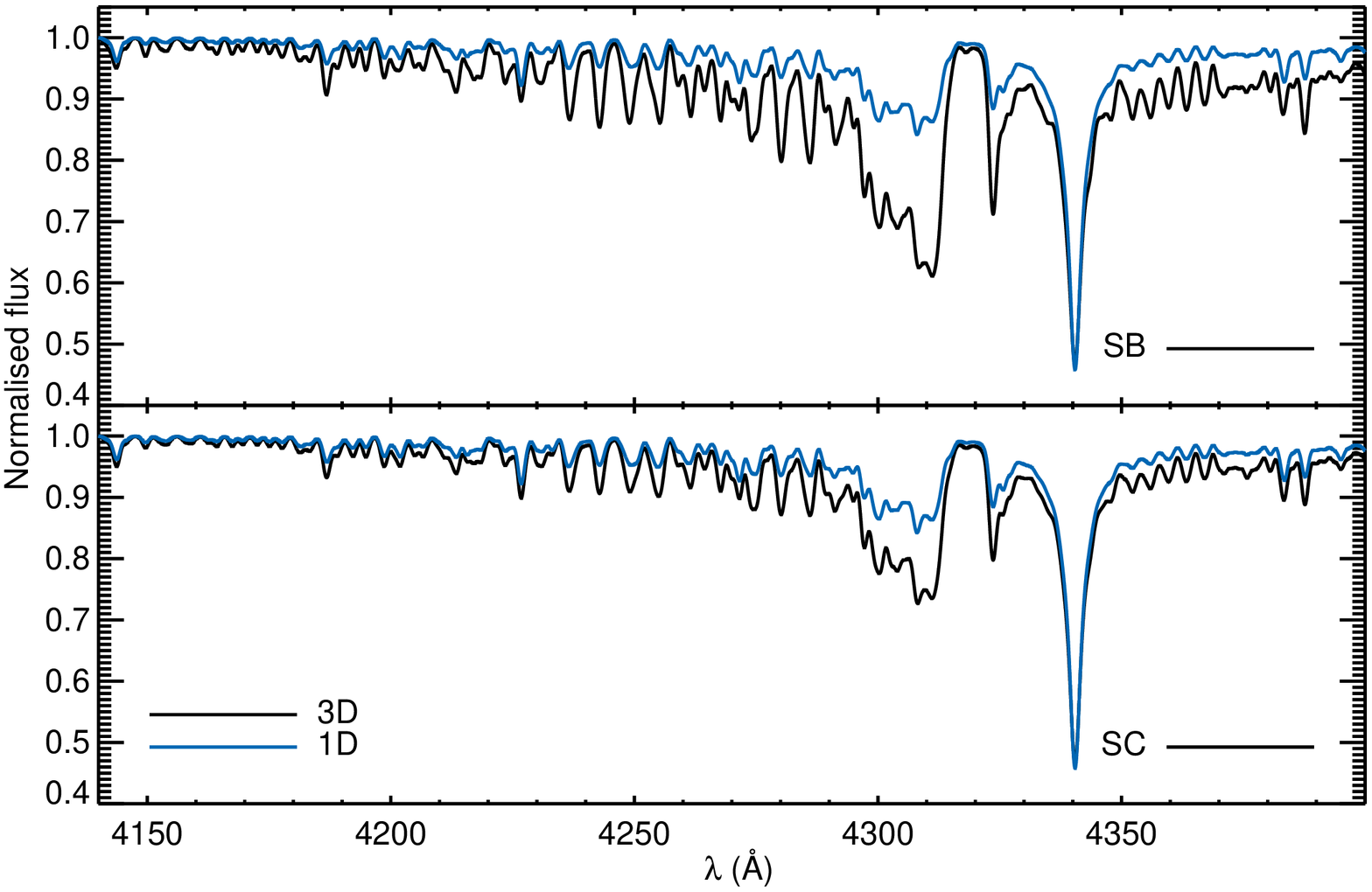} 
 \includegraphics[width=6.7cm]{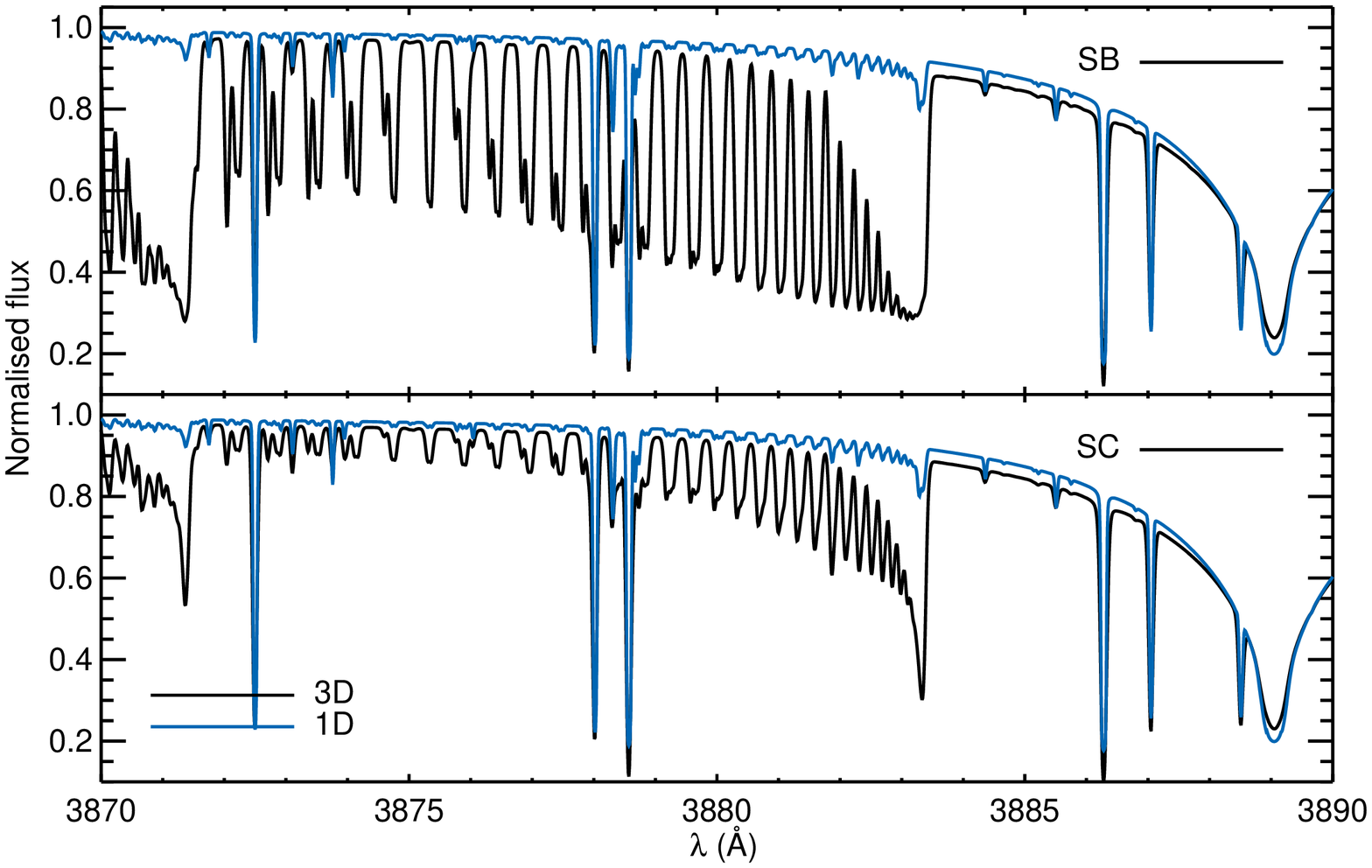} 
 \caption{Comparison of 3D (black) and 1D (blue) synthetic spectra of the G-band
 (left, degraded to SDSS-like spectral resolution of 150\,km/s) and of the 
CN-band (right, full spectral resolution), based on chemically consistent 
models B (top) and C (bottom).}
\label{fig1}
\end{center}
\end{figure}

\section{Results}

Given identical chemical composition, synthetic spectra of the molecular
bands based on 3D metal-poor model atmospheres are found to be significantly 
stronger than the corresponding 1D spectra (see Fig.\,\ref{fig1}), as 
previously pointed out by \cite{Bonifacio2013}.
By fitting the 3D synthetic G-band spectra with 1D spectra, varying the carbon
abundance A(C) at fixed C/O ratio, we derive 3D LTE abundance corrections
for carbon: $\Delta_{\rm 3D-1D}=$A(C)$_{\rm 3D} - $A(C)$_{\rm 1D}$.
The abundance corrections for oxygen are obtained by fitting
the 3D OH features with 1D spectra of different A(O). The results are 
summarized in the right part of Table\,\ref{tab1}. Inconsistencies in 
the chemistry (in particular C/O) between model atmosphere and 
line formation calculations (cases {\bf SAB}, {\bf SAC}) can lead to 
significant errors.
Further details can be found in 
\cite[Gallagher et al.\ (2016, 2017)]{Gallagher2016,Gallagher2017}.

\section{Implications}
We found substantial 3D effects on the formation of molecular
bands in a typical CEMP star, implying downward abundance
corrections for carbon and oxygen with respect to 1D results by
up to $-0.5$\,dex for the CH G-band and $-0.7$\,dex for the OH band.
Corrections for the CN-band could be even larger (work in
progress). The influence of non-LTE effects and the existence of
a stellar chromosphere may potentially change this conclusion.

\end{document}